\documentclass[letterpaper, 10 pt, conference]{ieeeconf}  

\IEEEoverridecommandlockouts                              

\usepackage{graphics} 
\usepackage{epsfig} 
\usepackage{mathptmx} 
\usepackage{times} 
\usepackage{amsmath} 
\usepackage{amssymb}  
\usepackage{cite}
\usepackage{epstopdf}
\usepackage{float}
\usepackage{caption}
\usepackage{subcaption}
\usepackage{url}

\title{\LARGE \bf
Sequential Monte Carlo Filtering Estimation of Ebola Progression in West Africa
}

\author {Narges~Montazeri~Shahtori{$^{1,*}$},
 Caterina~Scoglio  $^{1}$,
 Arash~Pourhabib  $^{2}$,
and~Faryad~Darabi~Sahneh$^{1}$
\thanks{This material is based upon work supported by the National Science Foundation under Grant No. SCH:1513639.}
\thanks{$^{1}$Shahtori, Sahneh, and Scolgio are with the Department of Electrical and Computer Engineering, Kansas State University, Manhattan,
KS, 66506. E-mail: \{nargesmsh92,caterina,faryad\}@ksu.edu.}
\thanks{$^{2}$Pourbabib is with the School of Industrial Engineering and Management, Oklahoma State University, Stillwater, OK 74078. E-mail: arash.pourhabib@okstate.edu.}}

\begin{document}

\maketitle
$ $
\thispagestyle{empty}
\pagestyle{empty}

\begin{abstract}
We use a multivariate formulation of sequential Monte Carlo filter that utilizes mechanistic models for Ebola virus propagation and available incidence data to simultaneously estimate the disease progression states and the model parameters. This method has the advantage of performing the inference online as the new data becomes available and estimates the evolution of basic reproductive ratio $R_0(t)$ of the Ebola outbreak through time. Our analysis identifies a peak in the basic reproductive ratio close to the time when Ebola cases were reported in Europe and the USA.

\end{abstract}

\section{Introduction}
Since December 2013, West Africa has experienced the largest Ebola outbreak with more than 20,000 infected cases reported \cite{nishiura2014early}. Secondary infections have also been reported in Spain and the United state \cite{WHO}. Ebola hemorrhagic fever is considered a highly infectious and lethal disease, raising serious concerns about the public health globally. Efforts to stop the spread of Ebola virus included intervention strategies such as surveillance, quarantining suspected cases and education of hospital workers in contact with Ebola patients \cite{chowell2004basic}. Side by side of these efforts, mathematical and computational epidemic models were developed and implemented with the aim of predicting newly infected cases as well as evaluating mitigation strategies.

The basic reproductive ratio, $R_{0}$, is one of the most relevant descriptors that helps health--care authorities to have a quantitative understanding of the severity of the disease outbreak and its time projection. The most common definition of basic reproductive number is the expected number of secondary infections produced by a typical single infection in a entirely susceptible population \cite{he2009plug}. Early estimation of the basic reproductive number and other relevant descriptors is crucial. Unfortunately, during early times available incidence data is severely scarce, making reliable inference extremely challenging.

Researchers have attempted to analyze the recent Ebola outbreak data and estimate the basic reproductive ratio. Althaus \cite{althaus2014estimating} used an offline optimization algorithm to find parameters of the susceptible-exposed-infected-susceptible (SEIR) epidemic model that fits best to collected Ebola data during a fixed time period. The major shortcoming of such approaches is that they provide an offline inference of an outbreak that is inherently dynamic and parameters of model change during disease evolution, so we need to keep tracking parameters when new data become available. Furthermore, since lots of factors such as intervention strategies could affect on parameters, we expect that the basic reproductive ratio changes during the disease evolution. Therefore we need techniques that are able to trace new data as they become available. Towevers \textit{et al.} \cite{towers2014temporal} estimated the basic reproduction ratio, $R_0(t)$,  by fitting exponential regression models to small successive time intervals of the Ebola outbreak. Therefore, they obtained an estimate of temporal variations of the growth rate. Their application of regression models ignores the systemic epidemiological information of Ebola progression---as reflected in the SEIR model--- and thus are more suitable for exploratory analysis of the incidence data. A more robust analysis should take advantage of our epidemiological knowledge of the dynamical system under study. Furthermore, the scarcity of incidence data during short time intervals impacts the stability of their regression model fitting.

In this paper, we estimate the states of the Ebola propagation and make inference about the basic reproductive ratio through time. To this end, we implement a sequential Monte Carlo (SMC) filter, an online inference method that allows simultaneous state and parameter estimation with improved accuracy as new streaming data becomes available. Sequential Monte Carlo filter is particularly powerful for inference about epidemic models which are inherently nonlinear and involve numerous uncertainties. Specifically, our SMC setting allows simultaneous estimation of the number of individuals at different infection stages as well as the parameters of our mechanistic epidemic model, providing posterior distributions of interest. In SMC, the distribution of interest is estimated by a large number of $N \gg 1$ random samples termed particles conditioned on the observations. A sampling mechanism propagates these particles \cite{kantas2009overview}. Afterward, we use the estimated values of the Ebola epidemic model parameters to determine the value of $R_0(t)$.

Compared with existing studies on the recent Ebola epidemics in West Africa, our approach has the advantage of performing the inference online as the new data becomes available and estimates the evolution of basic reproductive rate $R_0(t)$ of the Ebola outbreak through time. Interestingly, our analysis identifies a peak in the basic reproductive ratio close to the time when cases were reported in Europe and the USA.

The remainder of this paper is organized as follows. Section II presents basic tools used in sequential Monte Carlo filter and discusses the problem background in epidemic modeling. Section III outlines our modified SEIR model for Ebola and explains the particle setup and data, and Section IV presents main results of this study. Section V concludes the paper by suggestions for future research.
   
\section{Background}
\subsection{Epidemic Modeling}
Mathematical models of infectious disease offer a mechanistic framework to describe and study the spread of infections within human and animal populations, providing deep insight into their dynamics and suggesting practical strategies to reduce the severity of epidemics\cite{pellis2015eight}. Here, we introduce a brief background discussing the susceptible-exposed-infected-recovered (SEIR) model which is compatible with our understanding of Ebola virus epidemiology.

In the SEIR model, individuals are assumed to be in one of these compartments: susceptible, exposed, infected and removed/recovered. In this model, when susceptible individuals have contact with an infected person, they enter into the exposed compartment (E) with rate $\beta I$. Homogeneity of the population and how people have contact with each others in host population is represented by a percentage factor $c$. After the incubation period of disease, mean of $1/ \lambda$, they enter into the infected compartment (I). Infectious individuals move to the recovered/removed compartment (R) at rate $\gamma$ \cite{Keeling2008Modeling}.  The compartmental SEIR model is \cite{Keeling2008Modeling}

\begin{equation}\label{eq1}
\begin{split}
\frac{dS}{dt} &= - \beta cSI,\\
\frac{dE}{dt} &= \beta cSI - \lambda E,\\ 
\frac{dI}{dt} &= \lambda E - \gamma I,\\
\frac{dR}{dt} &= \gamma I.
\end{split}
\end{equation}

In this compartmental SEIR, the size of host population is assumed to remain constant throughout the evolution time, i.e., $ P = S + E + I +R $, and demographic effects are ignored.

An important mathematical tool in the study of epidemics is the basic reproductive ratio. Usually the basic reproductive number is defined as the expected number of secondary individuals produced by a typical single infected individual during its infectious period \cite{heffernan2005perspectives}. Thus, $R_0$ is a dimensionless value that represents the average number of additional susceptible people to whom an infected person passes the disease before he/she recovers \cite{newman2010networks}. For instance, if an infectious person passes the disease on three others on average, during his/her infectious lifetime, then $R_0 = 3>1$, indicating that the number of new infectious individuals would increase with each generation, so we can expect to experience an epidemic. Conversely, if $R_0 < 1 $ the disease will die out \cite{newman2010networks}. Thus, the basic reproductive ratio is a threshold condition for epidemics as $R_0=1$ separates the increments or decrements of new infected \cite{newman2010networks}. A common definition of $R_0$ in mathematical epidemiology is \emph{the average number of expected new infections over all possible infected types} \cite{diekmann1990definition}. A widely used technique for finding $R_0$, which is based on this  definition, is the \emph{next generation matrix} method \cite{diekmann1990definition}. Applying this technique to the SEIR model above finds $R_0 = \frac{\beta c}{\gamma}$ \cite{heffernan2005perspectives}.
  
\subsection{Sequential Monte Carlo Filter }
Sequential Monte Carlo (SMC) --- or particle filter --- refers to a class of statistical techniques that estimate unknown parameters, namely states in this context, as new streaming noisy observations becomes available \cite{sheinson2014comparison}. In SMC, we iteratively sample from the posterior distribution of parameters until the parameters converge to stationary values \cite{xue2014data}. This iterative sampling is updated using a stream of data, and as such, it enables us to modify our best guesses for the states according to actual observations. In the following, we explain the dynamic state--space model and estimation of posterior PDF briefly for the particle filters algorithm.
\paragraph{Dynamic state--space model}
The state--space model assumes the Markov property, i.e.,
\begin{equation}\label{eq3}
Pr(x_{k}|x_{0: k-1})= Pr(x_{k}|x_{k-1}).
\end{equation} 
and describes the distribution of the system state in the next step, as well as the observation, given the current state of the system. 
 More rigorously, a state--space model is defined as \cite{sheinson2014comparison,doucet2001introduction}: 
\begin{equation}\label{eq2}
\begin{split}
x_{k} &\sim f(x_{k}|x_{k-1}, \theta) ,\\ 
y_k &\sim f(y_k|x_{k},\theta),
\end{split}
\end{equation}
where $y_k$ represents $k$th observation, $x_{k}$ represents states corresponding to the $k$th observation, and $\theta$ represents parameters of the model. Therefore, $y_k$ depends only on $x_{k}$ and $\theta$ and $x_{k}$ depends on $x_{k-1}$ and $\theta$.
\paragraph{Estimation of posterior PDF}
Given the observation data $y_{1:k}$ up to time $k$, the ultimate goal is to define the  posterior distribution, $p(x_{k}, \theta|y_k)$, which describes the hidden state $x_{k}$ and parameters $\theta$ of the dynamical system. The estimation of posterior probability density function (PDF) based on Bayes' theorem is \cite{sheinson2014comparison}:
\begin{equation} \label{eq4}
\begin{split}
x_{k}, \theta_{k}| y_{1:k} &\propto f(y_k|x_{k},\theta) p_0(x_{k}, \theta)\\ &\propto f(y_k|x_{k},\theta) p_0(x_{k}, \theta|y_{1:k-1}).
\end{split}
\end{equation}
 
The sequential Monte Carlo filter approximates the posterior PDF as \cite{sheinson2014comparison}:
\begin{equation}\label{eq5}
Pr(x_{k} , \theta_{k} | y_{1:k})\approx \sum\limits_{i=1}^J w_k ^{(i)} \textbf{1}_{\lbrace(x_{k}, \theta_{k}) = (x_{k} ^ {(i)}, \theta^{(i)}_k)\rbrace}.
\end{equation}
where $\textbf{1}_{\lbrace . \rbrace}$ is the indicator function, $x_{k}^{(i)}$ is a particle and $w_k^{(i)}$ is its weight. The approximation is more accurate if the number of samples, i.e., particles, is large. Particle' weights are normalized thus \cite{skvortsov2012monitoring}
\begin{equation}\label{eq6}
\int Pr(x_{k} , \theta_{k} | y_{1:k}) dx_k \approx \sum\limits_{i=1}^J w_k^{(i)}.
\end{equation}
Among particle filter techniques are the bootstrap filter, auxiliary particle filter, and kernel density particle filter. In this paper, we use the latter, namely kernel density particle filter, due to its flexibility in modeling of non-linear processes as explained in Section \ref{sec:PF}.
 
\section{SMC for Ebola Epidemics}
In this paper, we use the discrete-time SEIR model which is compatible with the epidemiology of the Ebola virus. 
\subsection{Modeling of Ebola} 
 The state variables $S_t$, $E_t$, $I_t$, and $R_t$ denote the fraction of people who are susceptible, exposed, infected and recovered or removed, at time step $t$, respectively which $t$ step is one day. For our analysis, we use the discrete--time form of equation \ref{eq1} with stochastic fluctuation and the following assumptions to modify the original SEIR model \ref{eq1}.

\textbf{Assumption 1:} Since the population is much greater than the number of infected cases of Ebola, we assume $S \simeq 1.$ Therefore, the equation for evolution of $E(t)$ in (1) simplifies to:
\begin{equation}\label{eq7}
E_{t+1} = E_{t} + \beta c_t I_t - \lambda E_t. 
\end{equation}

\textbf{Assumption 2:} We assume that $c_t$, representing how the population is mixed, is dynamic due to different intervention strategies to prevent the spread of Ebola such as social distancing and quarantining. Specifically, we assume $c_t$ is decreasing at rate $\alpha$, i.e.,
\begin{equation}\label{eq8}
c_{t+1}= c_t -\alpha c_t.
\end{equation}
This is a simplified assumption to account for different intervention strategies. We assumed that when the control measured are introduced and information regarding Ebola disease is disseminated, the transmission rate decays exponentially.

According to above assumptions and modifications to the SEIR model \ref{eq1}, we propose the following set of stochastic difference equations as our base epidemic model for the Ebola spread.
\begin{equation}\label{eq9}
\begin{split}
c_{t+1} &=c_t - \alpha c_{t} + \xi_{\alpha}, \\
E_{t+1} &=E_t + \beta c_{t} I_{t}- \lambda E_{t}+ \xi_{\lambda} - \xi_{\beta}, \\
I_{t+1} &= I_t + \lambda E_{t} - \gamma I_{t} - \xi_{\lambda} +  \xi_{\gamma},\\
R_{t+1} &= R_t + \gamma I_{t}- \xi_{\gamma}\\
D_{t+1}  &= \varphi R_{t+1} = \varphi R_t + \varphi \gamma  I_{t}- \xi_{\gamma} + \xi_{\varphi}.
\end{split}
\end{equation}

In the above equations, $\xi_{\chi}$ where $\chi\in\lbrace \alpha, \beta, \lambda, \gamma, \varphi \rbrace$ is a random component, with zero mean and given variance $\sqrt{\chi}/P$, where $P$ is the population size. The variance of noises are due to stochasticity of the underlying process \cite{sheinson2014comparison}. Each of these component are assumed to be uncorrelated. 

\subsection{Filtering Setup}\label{sec:PF}

The technique of bootstrap filter and auxiliary particle filter are employed to generate the kernel density particle filter method. In bootstrap filter, probability density $p(x)$ is estimated by a set of particles and at each round their weights are computed and those particles with small weights, are eliminated. After each round, the surviving particles produce new particles.
 The main problem  when bootstrap filter is employed is that $ w_k^{(i)}$ might become very small for some particles and affect the accuracy of the particle filter. A proper way to decrease the number of particles is to select the importance probability function close to the optimal one \cite{doucet2001introduction}. Auxiliary particle filter aims is predicting which particles will have a small weight and minimizes those particles with small weights. With respect to these two filters, the kernel particle filter not only minimizes those kinds of particles with small weights, but also estimates the unknown parameters \cite{chang2005kernel}. In this method, the main goal is to reduce the mean integrated square error of the kernel approximation and the posterior PDF in each step.

 Using the kernel density, we approximate and reproduce parameters of the system in addition to the posterior probability density function, $p(x_{t_k},\theta _{k}|y_{1:k})$, of $x_{t_k}$. Here, for each particle $i$, $\theta^{(i)}_{k}$ is the value of parameters at time $t$ which we only have $k$ observations up to time $t$.

We choose reported Ebola data in Guinea, one of the three major West Africa countries that experienced the Ebola outbreak and we analyze the cumulative cases and death counts. To update parameters and states using kernel particle filter, we need data. However, the number of cases and death in Guinea are reported at irregular intervals. To address this issue, we implement particle filter only for those time intervals that we have information and use the last updated parameters and states that are generated by particle filter to update states for those intervals that we do not have any information. 
 \subsubsection{Evolution Setup}
According to our Ebola model in \ref{eq9}, we can write the state--space model required for SMC method as
\begin{equation}\label{eq10}
x_{t+1}|x_{t} \sim N_\Omega (g(x_{t} , \theta),Q(\theta)), 
\end{equation}
where $ x_{t}=[c_{t}, E_{t}, I_{t}, R_{t}, D_{t}]^T$ is the state and
\begin{equation}\label{eq11}
\begin{split}
g(x_{t},\theta) &=
    \begin{bmatrix}
    c_{t}- \alpha c_{t} \\
    E_{t} + \beta c_{t} I_{t} - \lambda E_{t}\\
     I_{t} + \lambda E_{t} - \gamma I_{t}\\
     R_{t}+\gamma I_{t}\\
      \varphi I_{t} + \varphi \gamma  I_{{t}}\\
    \end{bmatrix},
\\
Q(\theta) &= \dfrac{1}{P^2}
    \begin{bmatrix}
    \alpha & 0& 0 & 0 & 0 \\
     0 & \lambda+\beta & - \lambda & 0 & 0 \\
     0 & - \lambda & \lambda+\gamma & -\gamma & -\gamma\varphi \\
     0 & 0 & - \gamma &  \gamma & \gamma\varphi \\
      0& 0 & -\gamma\varphi & \gamma\varphi & \gamma\varphi^2
    \end{bmatrix}.
\end{split}
\end{equation}
where $Q(\theta)$ is the covariance matrix of $ x_{t}=[c_{t}, E_{t}, I_{t}, R_{t}, D_{t}]^T$ according to the state-space model (\ref{eq9}) with stochastic fluctuations. Here, $\mathcal{N}_{\Omega} (\mu, \Sigma)$ represents the truncated normal distribution where $\Omega=\lbrace(c_{t}, E_{t}, I_{t}, R_{t}, D_{t} ): c_{t}, E_{t}, I_{t}, R_{t}, D_{t}\geq 0, E_{t}+ I_{t}+ R_{t} \leq 1 \rbrace $.
\subsubsection{Observation Setup}
Observations are positive values and in each reported day, WHO reports cumulative cases and death. Using SEIR model, we only estimate number of infected and recovered individuals at each time step. Therefore, to configure SMC observation setup based on reported data, we need to estimate cumulative cases and death. To this end, using states $I$ and $R$ in equation \ref{eq9}, we model the observations $Y$ as follows \cite{sheinson2014comparison}:
\begin{equation}\label{eq12}
Y \sim \mathcal{N} (\mu_{Y}, \Sigma_Y),
\end{equation}
where 
\begin{equation}\label{eq13}
\begin{split}
Y &=
\begin{bmatrix}
log(y_{I,k})\\
log(y_{D,k})\\
\end{bmatrix}
\end{split}
\begin{split},
\mu_Y &=
\begin{bmatrix}
b_I (I_{t} + R_{t})^{\zeta _I}\\
b_D (D_{t})^{\zeta _D} \\
\end{bmatrix} 
\end{split},
\begin{split}
\Sigma_Y \sim
\begin{bmatrix}
\sigma_I & 0\\
0 & \sigma_D\\
\end{bmatrix} 
\end{split}.
\end{equation}

Index $D$ and $I$ in observation matrix $Y$, represent total number of dead and infected individuals at time $t$, as reported by WHO. In matrix $\mu_Y$, $I$ represents the estimation of number of infected individuals (state $I$) and $R$ represents estimation of state $R$; individuals who are dead or recovered. $\zeta s, \sigma s$ and $b$s for both infected and dead are typically unknown, but we can predict these values by linear regression methods\cite{skvortsov2012monitoring}. In particular, $b$s are multiplicative constants and $\zeta$s are  power-law exponents which can be calculated based on the significant of dispersed of data points. Since fluctuations exist in the data, $\zeta$s are not estimated precisely. The scaling law for $\Sigma_Y$, standard deviation, is $\sigma_. \propto \dfrac{1}{\sqrt{P}}$ where $P$ is the population size \cite{skvortsov2012monitoring}. 
\subsubsection{Kernel density particle filter} \label{algorithmsection}
Model (\ref{eq12}) and (\ref{eq13}) define the likelihood of observations, $y_k$, given $x_{t_k}$ and $\theta$ --- $p(y_k|x_{t_k},\theta)$ --- as a log normal distribution with mean $\mu_Y,$ which is two-by-one vector and the standard deviation, $\Sigma_Y$, which is a two-by-two diagonal matrix with diagonal element equal to $\sigma_I$ and $\sigma_D$.  
 Initially, particle $i$ which $i\in{\lbrace1,...,J\rbrace}$ is sampled from an initial probability density function (PDF) $ p(x_0)$. For time step $k = 1$, weights are equal to $1/J$ for all particles, and $\theta_0$ and $x_0$ are generated by random sampling from specific distributions. Suppose $k+1$th observation becomes available. The following steps present algorithm which applied kernel particle filter, to update and estimate $p(x_{t_{k+1}},\theta|y_{1:k+1})$ for $ k > 1$ \cite{sheinson2014comparison}. 
 \begin{enumerate} 
 \item \label{require.step1}\textbf{Calculate $m_{k+1}^{(i)}$:}  $m_{k+1}^{(i)}= a \theta _{k}^{(i)}+(1-a) \bar {\theta_{k}}$. 
 
 \item \textbf{Compute the expectation of $x_{t_{k+1}}^{(i)}$: } 
 
 $\mu_{k+1}^{(i)}= E(x_{t_{k+1}}|\theta_{k}^{(i)},x_{t_k}^{(i)})$, for all $i\in\lbrace 1, 2,.., J\rbrace$. 
 \item \textbf{Compute auxiliary weights and normalize them:} 
 
 $g _{k+1}^{(i)}= w_{k}^{(i)} p(y_{k+1}|\mu_{k+1}^{(i)},m_{k+1}^{(i)})$ , $g_{k+1}^{(i)}= \frac{g_{k+1}^{(i)}}{\sum\limits_ {l=1}^{J} g_{k+1}^{(l)}}$.
 \item \textbf{Sampling:} Select an index $j$ randomly and afterwards sample $x^{(j)}_{k}$ with its weights $\lbrace g_{k+1}^{(1)},...,g_{k+1}^{(J)} \rbrace$.  
\item \textbf{Reproduce the parameters:}
$\theta _{{k+1}}^{(i)}\sim \mathcal{N_\omega}(m_{k+1}^{(j)},V^{\theta}_{k+1})$,
where $\mathcal{N_\omega}(\mu,\sigma)$ is a truncated normal distribution. 

\item \textbf{Sample the $x_{t_{k+1}}^{(i)}$:}
$x_{t_{k+1}}^{(i)} \sim p(x_{t_{k+1}}^{(i)}|\theta_{{k+1}}^{(i)},x_{t_k}^{(j)})$, for  all $i\in\lbrace 1, 2,.., J\rbrace$.

\item \textbf{Recompute the expectation of $x_{t_{k+1}}^{(i)}$:}

 $\mu_{k+1}^{(i)}= E(x_{t_{k+1}}^{(i)}|\theta_{k}^{(i)}, x_{t_k}^{(j)})$, for all $i\in\lbrace 1, 2,.., J\rbrace$.

 \item \label{require.step2}\textbf{Compute weights and normalize them again:} 

 $w_{k+1}^{(i)}= \frac {p(y_{k+1}|x_{t_{k+1}}^{(i)}, \theta_{{k+1}}^{(i)})} {p(y_{k+1}|\mu_{k+1}^{(j)}, m_{k+1}^{(j)})}$ , $w_{k+1}^{(i)}= \frac{w_{k+1}^{(i)}}{\sum\limits_ {l=1}^{J} w_{k+1}^{(l)}}$.
 
\end{enumerate}
In above, $a= 1-h^2$ and $h= 1-(\frac{3\phi-1}{2\phi})^2$. These two quantities control the smoothness of kernel density estimation, while $\phi \in (0,1)$ is a discount factor which reduces the chance of failure in the filter. Readers are encouraged to refer to \cite{sheinson2014comparison} for more details. Typically, $\phi$ is assumed to be a number between $.95$ and $.99$. 

\subsection{Data Explanation}
\begin{figure}[h]
    \centering
    \includegraphics[width=0.5\textwidth]{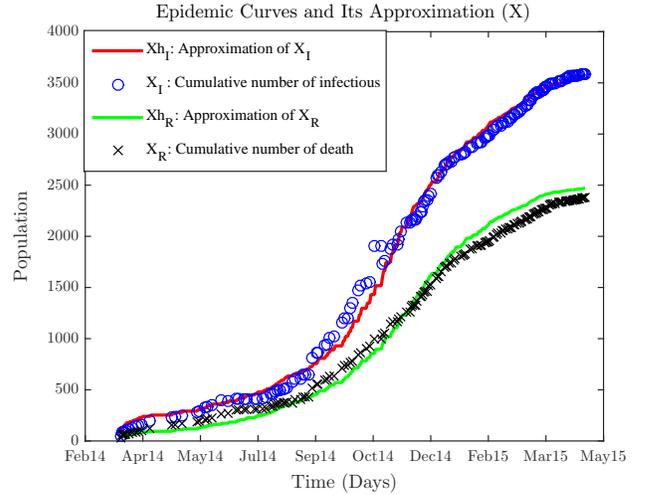}
    \caption{Cumulative cases and death data and their estimation by the particle filter in Guinea, reported by [WHO]}
    \label{figure.1}
\end{figure}
The Ebola virus, commonly known as Ebola, causes a serious illness which is  fatal if untreated in most cases \cite{WHO}. It is transmitted via direct contact with blood, secretions, organs, or other bodily fluids of infected individuals \cite{lekone2006statistical}. The incubation period, the time interval from infection with Ebola virus to the onset of symptoms, is between two to twenty one days \cite{WHO}. First symptoms of Ebola include the sudden onset of fever, fatigue, muscle pain, headache, and sore throat \cite{WHO}. Since approximately December 2013, West Africa has been affected by this virus. However, The World Health Organization (WHO) declared the epidemic to be a public health emergency of international concern on August 8, 2014 \cite{WHO}. We analyze the cumulative case and death counts in Guinea, one of the three West Africa Countries that experienced the Ebola outbreak. Cumulative cases are classified into three categories: confirmed, probable, suspected cases. Similarly, we have three cases for death counts. Confirmed cases are those individuals who are diagnosed by polymerase chain reaction (PCR) method. On the other hand, suspected and probable cases denote those individuals that have symptoms of Ebola but it is not confirmed if they are actually infected \cite{nishiura2014early}. We should mention that the cases were reported at irregular intervals. This data has been collected from reports of WHO available at \url{http://www.healthmap.org/ebola/}.
\section{Results}
 We estimate the states of Ebola propagation from March 23, 2014 to April 30, 2015 having data reported only in $T=170$ days. Guinea population in 2015 was estimated to be around 12,500,000, however, the population in danger to be infected by the Ebola virus was estimated to be roughly about $P=1,000,000$. 
  
 We estimate parameters by sampling from a log-normal distribution using $J= 5,000$, number of particles. The sensitivity to changes in discount factor is chosen as $\phi=0.95$ and initial weights are all set equal to $w_0=J^{-1}$. To run the particle filter, the initial state, $x_0$, and initial parameters, $\theta_0$, are generated randomly. Specification of initial priors distributions are reported in Table \ref{para} and \ref{states} .
 
\begin{table}[h]
\caption{Priors specification for parameters} 
\begin{tabular*}{\linewidth}{@{\extracolsep{\fill}}llr}
\hline
Parameters& Priors\\
\hline
 mitigation rate ($\alpha$) &$\mathcal{U}(.0059,.00593)$  \\
transmission rate ($\beta$ )  &$\mathcal{U}(.259,.379)$         \\
latency rate ($\lambda$)& {Beta}$(78,577)$             \\
 recovery/remove rate ($\gamma$)& {Beta}$(21,246)$              \\
 fatality rate ($\varphi$)& {Beta}$(37,15)$             \\
\label{para}
\end{tabular*}
 \end{table}
 \begin{table}
\caption{Priors specification for states} 
\begin{tabular*}{\linewidth}{@{\extracolsep{\fill}}llr}
\hline
States   & Priors \\
\hline
 $c$     &$\mathcal{U}(.36,.40)$\\
 $E$&$\mathcal{U}(.000128,.000141)$\\
 $I$& $\mathcal{U}(.000050,.000061)$\\
$R$&$\mathcal{U}(.000042,.000058)$\\
$D$&$\mathcal{U}(.000029,.000030)$
\label{states}                        
\end{tabular*}
\end{table}
Based on collected data and expert opinion, some measurements for $\gamma$, $\lambda$, and $\varphi$ are available. Therefore, we specify beta distributions for each of parameters, using \textit{Beta buster} and \textit{BetaSlicer} \cite{jones2014prior,chunbetabuster}. Based on collected information, the average incubation period,$\lambda ^{-1}$ is less than $21$ days with $95\%$ confidence interval and mean around $8$ days \cite{WHO}. Fatality rate, $\varphi$, is less than $80\%$ with $95\%$ confidence interval and mode $71\%$ \cite{WHO,chowell2014transmission,team2014ebola}. The average duration of illness onset to death or recovery, $\gamma^{-1}$, is around 12 days \cite{chowell2014transmission,team2014ebola}. Since we do not have enough information about transmission and mitigation rates, $\beta$ and $c$, we assume uniform distributions.

Since we do not have enough information about initial states, we use uniform distributions. For observation constant in SMC filter, we assume that $b_I = .88$ and $b_D = .54$ and standard deviation, $\Sigma_Y$ is a two by two diagonal matrix whose diagonal elements are $\sigma_I = .00125$ and  $\sigma_D = .00085$. Power -- law constant $\zeta$, for infected individuals, is $.88$ and for dead individuals is $.68$. 
\begin{figure}
    \centering
    \begin{subfigure}[b]{0.5\textwidth}
        \centering
        \includegraphics[width=\textwidth]{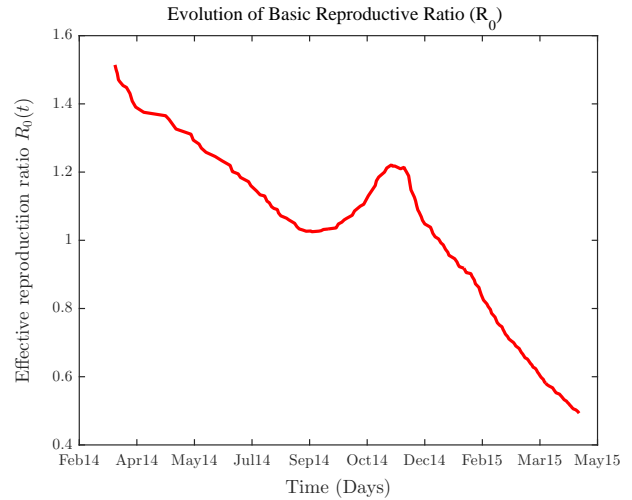}
        \caption{Changes in basic reproductive ratio over disease evolution}
        \label{figure.R_0}
    \end{subfigure}
    \hfill
    \begin{subfigure}[b]{0.5\textwidth}
        \centering
        \includegraphics[width=\textwidth]{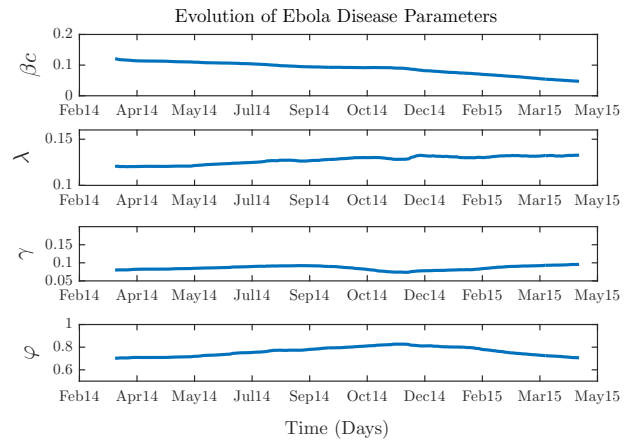}
        \caption{Parameters changes over Ebola disease evolution }
        \label{figure.para}
    \end{subfigure}
    \hfill
    \caption{Variation in $R_0$ and parameters during disease evolution}
    \label{fig.para_R}
\end{figure}
Fig. \ref{figure.1} shows cumulative cases and deaths data and their estimation by the particle filter in Guinea. In our model, the basic reproductive ratio is equal to $R_0 (t)= c \beta \gamma^{-1}$. The result indicates that transmission rate, latency rate and recovery rate are not constant during the disease evolution. Therefore, as demonstrated in Fig. \ref{figure.R_0}, $R_0 (t)$ is not constant neither. 

The maximum value of the basic reproductive number is $1.51$ on March 2014 and it decreases until September 2014 which $R_0 \sim 1$. Afterward, a pick is occurred on October 2014 and after that it decreases. We can see in Fig. \ref{figure.para} that transmission rates change during the disease evolution. In \cite{gomes2014assessing}, $R_0(t)$ is estimated as a single value with confidence interval, while in \cite{towers2014temporal} for the reproduction ratio, $R_0(t)$, is computed by fitting exponential growth curves to small successive time intervals of the Ebola outbreak. Instead, our method finds the basic reproductive ratio, $R_0(t)$, as a continuous function of time during the Ebola evolution and for each time a probability function is represented. Using this method, we can also see that not only $R_0(t)$ changes over time, but also parameters such as $\beta$, $\lambda$ and $\gamma$ change. 
\section{Discussion and conclusion}
Our analysis identifies a correlation between $R_0(t)$ temporal variations and important events in the 2014 Ebola outbreak. For instance, a reduction of $R_0(t)$ can be seen around the time WHO first announced the Ebola outbreak in Guinea. This reduction can be explained by taking into account the introduction of some initial medical support and public awareness. Conversely, a peak of $R_0(t)$  corresponds to the first Ebola cases in Europe and the USA.

Further improvement of our method would involve accounting for intermittent measurements and heterogeneous variance in the particle filtering scheme. Furthermore, our experience with numerical simulations showed fairly consistent outcomes. However, a more objective quantification of involved uncertainties can be a great addition to the current work. At the end, we would like to reiterate that we did not account for any spatial dependencies in our analysis. A spatial implementation of the particle filter can account for several countries in West Africa all at once.
\section*{Acknowledgment}
The authors would like to thank Joan Salda{\~n}a for fruitful discussions regarding basic reproductive ratio in realistic epidemics, Nora Bello for fruitful discussions and suggestion regarding the priors specifications, and Ala Fard for organizing Ebola incidence data for our analysis.

\bibliographystyle{IEEEtran}
\bibliography{EKF_bib}

\end{document}